\documentclass{article}
\usepackage{float}

  \usepackage[creativeai, final]{neurips_2025}

\usepackage[utf8]{inputenc} 
\usepackage[T1]{fontenc}    
\usepackage{hyperref}       
\usepackage{url}            
\usepackage{booktabs}       
\usepackage{amsfonts}       
\usepackage{nicefrac}       
\usepackage{microtype}      
\usepackage{xcolor}         
\usepackage{graphicx}
\usepackage{amsmath}
\usepackage{amssymb}
\usepackage{algorithm}
\usepackage{algorithmic}

\title{ShrutiSense: Microtonal Modeling and Correction in Indian Classical Music
}

\author{%
  Rajarshi Ghosh\thanks{Equal contribution} \\
  Lone Star College\\
  \And
  Jayanth Athipatla\footnotemark[1] \\
  University of Nebraska at Omaha \\
}

\begin{document}

\maketitle

\begin{abstract}
Indian classical music relies on a sophisticated microtonal system of 22 \emph{Shrutis} (pitch intervals), which provides expressive nuance beyond the 12-tone equal temperament system. Existing symbolic music processing tools fail to account for these microtonal distinctions and culturally-specific \emph{raga} grammars that govern melodic movement. We present \textbf{ShrutiSense}, a comprehensive symbolic pitch processing system designed for Indian classical music, addressing two critical tasks: (1) correcting westernized or corrupted pitch sequences, and (2) completing melodic sequences with missing values. Our approach employs complementary models for different tasks: a Shruti-aware Finite-State Transducer (FST) that performs contextual corrections within the 22-Shruti framework and a Grammar-Constrained Shruti Hidden Markov Model (GC-SHMM) that incorporates raga-specific transition rules for contextual completions. Comprehensive evaluation on simulated data across five ragas demonstrates that ShrutiSense (FST model) achieves 91.3\% Shruti classification accuracy for correction tasks, with example sequences showing 86.7--90.0\% accuracy at corruption levels of 0.2 to 0.4. The system exhibits robust performance under pitch noise up to \textpm50 cents, maintaining consistent accuracy across ragas (90.7--91.8\%), thus preserving the cultural authenticity of Indian classical music expression.
\end{abstract}

\section{Introduction}

Indian classical music represents one of the world's most sophisticated microtonal systems, employing 22 distinct pitch intervals called \emph{Shrutis} within each octave \citet{bharata_natyashastra}. Unlike Western music's 12-tone equal temperament, this system provides finer granular control over pitch relationships, enabling the subtle melodic ornamentations and emotional expressions that characterize ragas. The concept of Shruti, first formalized in Bharata's \emph{Natya Shastra} (circa 200 BCE--200 CE), divides the octave into 22 mathematically precise intervals, each serving specific melodic and emotional functions within different raga contexts.

\subsection{Problem Definition and Real-World Applications}

Processing Indian classical music poses distinct challenges for contemporary digital music technologies due to its microtonal structure and grammar-rich melodic design. These challenges appear across multiple real-world domains. In digital music education, online platforms and mobile applications need precise pitch feedback for learners practicing ragas, yet existing Western-oriented tools often miscorrect notes in ways that violate raga grammar. For archival digitization, historical recordings and manuscripts require symbolic transcription methods that retain cultural nuance, but most tools reduce pitch data to the 12-tone equal temperament (12-TET) system, erasing vital microtonal content. Performance assistance technologies, such as live accompaniment and tuning systems, demand real-time correction algorithms that honor microtonal aesthetics without compromising melodic flow. Similarly, music information retrieval platforms, such as search engines and recommendation systems, benefit from culturally-aware analyses capable of identifying raga-specific structures and pitch relationships. To support these domains, ShrutiSense introduces two key computational tasks: pitch correction, which transforms noisy or Western-biased pitch sequences into culturally appropriate output conforming to Shruti resolution and raga grammar; and melodic completion, which fills in missing pitches using contextual musical cues and embedded cultural knowledge to restore and enhance symbolic content for interactive composition, transcription, and preservation purposes.

\subsection{Technical Contributions}

Our primary contributions include: (1) formalization of the 22-Shruti system within a computational framework with explicit raga grammar encoding, (2) development of grammar-constrained probabilistic models that balance pitch accuracy with cultural authenticity, (3) comprehensive evaluation framework demonstrating superior performance over naive quantization approaches, and (4) open-source implementation\footnote{The code is available at \url{https://github.com/rajarshi51382/microtonal-carnatic}} enabling reproducible research in computational ethnomusicology.

\section{Mathematical Foundations}

\subsection{Shruti Theory and Scale Selection}

We formalize the 22-Shruti system as a logarithmic frequency division of the octave. Let $f_0$ denote the fundamental frequency of the tonic (\emph{shadja}). Following the theoretical framework established by \cite{bhatkhande1934comparative}, we define each Shruti $s_i$ as $f_i = f_0 \cdot 2^{c_i/1200}$ where $c_i$ represents the cent value of the $i$-th Shruti relative to the tonic. Our implementation uses the widely-accepted 22-Shruti scale with cent values:

\begin{multline}
\mathbf{C} = \{0, 90, 112, 182, 204, 294, 316, 386, 408, 498, 520, \\
590, 612, 702, 792, 814, 884, 906, 996, 1018, 1088, 1110\}
\end{multline}

This scale selection represents a consensus among contemporary musicologists and provides computational tractability while preserving essential microtonal distinctions. Alternative Shruti theories exist (e.g., Sarangadeva's 22-Shruti system, Ramamatya's modifications), but our chosen scale aligns with modern performance practice and digital applications.

The minimum distinguishable pitch interval is:
\begin{equation}
\delta_{\text{shruti}} = \min_{i,j} |c_i - c_j|, \quad i \neq j = 22 \text{ cents}
\end{equation}

\subsection{Raga Grammar Construction}

We model each raga as a directed graph $G = (S, T)$ where $S = \{s_1, s_2, \ldots, s_{22}\}$ is the set of Shruti positions and $T \subseteq S \times S$ represents permissible transitions between Shrutis. The raga grammar construction process involves classical treatises defining arohana (ascending) and avarohana (descending) scales, vadi-samvadi relationships, and varjya (forbidden) notes. Our current implementation covers five major ragas: Yaman, Bhairavi, Bilaval, Kalyan, and Khamaaj, representing diverse tonal structures and transition patterns. The grammar function $G(s_i, s_j)$ is defined as:

\begin{equation}
G(s_i, s_j) = \begin{cases}
1 & \text{if } (s_i, s_j) \in T \\
0 & \text{otherwise}
\end{cases}
\end{equation}

Transition weights incorporate both theoretical and empirical preferences:

\begin{equation}
w(s_i, s_j) = G(s_i, s_j) \cdot \exp(-\alpha \cdot d(s_i, s_j)) \cdot \text{pakad\_bonus}(s_i, s_j)
\end{equation}

where $d(s_i, s_j)$ represents interval distance, $\alpha = 0.1$ controls stepwise preference, and pakad\_bonus provides additional weight for characteristic phrase patterns.

\section{Data Collection and Preprocessing}

\subsection{Dataset Construction}
We synthesized 1,000 pitch sequences from canonical raga grammar rules (without model-generated sequences), then applied controlled corruption patterns simulating common digitization errors such as random note substitutions (10-50\% of sequence), and missing value patterns (10-50\% missing rates).

ShrutiSense accepts multiple input formats, including MIDI data (Note-on events converted to cent deviations from tonic), symbolic notation (manual pitch annotations in cents or Shruti labels), and audio-derived pitch (fundamental frequency tracks from audio analysis). The preprocessing pipeline normalizes all inputs to cent values relative to a detected or specified tonic, handles octave disambiguation, and segments long sequences into musically meaningful phrases (typically 8-32 notes).

\section{Grammar-Constrained Hidden Markov Model}

The Grammar-Constrained Shruti Hidden Markov Model (GC-SHMM) is defined over a state space $S = \{s_1, s_2, \ldots, s_N\}$, where $N$ represents the number of active Shrutis in a given raga, typically ranging from 7 to 10. The observation sequence $O = \{o_1, o_2, \ldots, o_T\}$ contains noisy pitch values measured in cents. Emission probabilities are modeled using Gaussian distributions centered around the theoretical frequencies of each Shruti, with variance $\sigma = 25$ cents reflecting empirical pitch deviation$$P(o_t | s_i) = \frac{1}{\sqrt{2\pi\sigma^2}} \exp\left(-\frac{(o_t - \mu_i)^2}{2\sigma^2}\right)$$where $\mu_i$ is the cent value of state $s_i$.

A key feature of GC-SHMM is its enforcement of raga grammar through constrained transitions. The probability of transitioning from state $s_i$ to $s_j$ is given by $P(s_j | s_i) = w(s_i, s_j)/Z_i$ if the pair satisfies the grammar function $G(s_i, s_j) = 1$, and zero otherwise, where $$Z_i = \sum_{k: G(s_i, s_k) = 1} w(s_i, s_k)$$ensures proper normalization. Additionally, direction-aware transition matrices $A_{\text{up}}$ and $A_{\text{down}}$ distinguish ascending and descending movements, with melodic direction inferred via pitch gradients.

For task-specific inference, the correction task utilizes the Viterbi algorithm to identify the most probable Shruti sequence, computed by maximizing over permissible transitions and emissions: $$\delta_t(j) = \max_{i: G(s_i, s_j) = 1} [\delta_{t-1}(i) \cdot P(s_j | s_i)] \cdot P(o_t | s_j)$$For completion tasks with missing pitches, the forward-backward algorithm is employed, where $\alpha_t(i)$ and $\beta_t(i)$ represent the forward and backward probabilities respectively, and the marginal probability $\gamma_t(i)$ is computed as $$\gamma_t(i) = \frac{\alpha_t(i) \beta_t(i)}{\sum_j \alpha_t(j) \beta_t(j)}$$At missing positions, a uniform emission $P(o_t | s_i) = 1$ is used, ensuring the model relies entirely on contextual grammar constraints to infer the correct Shruti.

\section{Shruti-Aware Finite-State Transducer}

The Shruti-aware Finite-State Transducer (FST), denoted as $\mathcal{T}$, maps input pitch sequences to corrected or completed outputs by applying weighted edit operations. The transducer supports four operation types: direct matches mapping input pitches to their nearest Shruti states, insertions to add missing Shruti symbols, deletions to remove spurious tokens, and substitutions to correct misclassified pitches. The total path cost combines three components: pitch fidelity, grammar compliance, and edit penalties, computed as 
\begin{equation}
    w(q_i, q_j, o, s) = \lambda_1 \cdot c_{\text{pitch}}(o, s) + \lambda_2 \cdot c_{\text{grammar}}(s_{\text{prev}}, s) + \lambda_3 \cdot c_{\text{edit}}
\end{equation}
where $c_{\text{pitch}}(o, s) = -|o - \mu_s|/50$ represents the normalized pitch distance, $c_{\text{grammar}}(s_{\text{prev}}, s)$ evaluates raga transition validity and likelihood, and $c_{\text{edit}}$ applies fixed penalties for non-match operations. In cases where transitions violate raga rules ($G(s_{\text{prev}}, s) = 0$), grammar cost is set to $-\infty$ to prohibit the path. We choose the parameters $\lambda_1 = 0.6$, $\lambda_2 = 0.3$, and $\lambda_3 = 0.1$, to ensure that pitch fidelity is emphasized without compromising grammatical integrity.

For context-aware completion, the FST leverages a sliding window of three surrounding notes to enhance prediction accuracy. The scoring function $\text{score}(s_{\text{candidate}}) = w_{\text{base}} + w_{\text{transition}} + w_{\text{pakad}} + w_{\text{position}}$ integrates multiple musical features: the base raga membership of the candidate note, its transition probability with contextual neighbors, alignment with known pakad motifs, and its positional relevance within the tala cycle—especially when the candidate occupies a metrically significant location such as vadi or samvadi on a strong beat. This holistic weighting strategy enables informed and musically meaningful completions in performance scenarios where input is sparse or noisy.

\section{Implementation Details}

The GC-SHMM algorithm operates with a time complexity of $O(TN^2)$, where $T$ denotes the sequence length and $N$ the number of active states, and a space complexity of $O(TN)$. In contrast, the Finite-State Transducer (FST) approach exhibits a time complexity of $O(TM^2)$ with lattice size $M$, typically set to 23 to account for 22 Shrutis plus epsilon transitions. An analysis of common failure cases reveals several limitations in the current implementation. First, ornament boundaries such as rapid microtonal oscillations (e.g., gamak, meend) pose challenges due to the system’s assumption of note-level granularity. Second, cross-raga modulation—where melodic sequences blend characteristics from multiple ragas—exceeds the model's capacity under a single-grammar assumption. Third, extreme pitch deviations over \textpm75 cents often result in incorrect Shruti classification, particularly in noisy input conditions. Finally, context sparsity in completion tasks—such as missing values at sequence boundaries or in contiguous gaps of five or more notes—significantly reduces prediction accuracy. We hope these observed limitations help guide future development, refinement of algorithmic strategies, and best practices for effective system use.

\subsection{Preprocessing}

We focus on the correction aspect of ShrutiSense, as it's the most applicable real-world use; the completion task is included only to test model capability.
The preprocessing pipeline extracts microtonal pitch from a WAV file and represents it as a sequence of cents. Using Librosa, we load the audio and retrieve its time series and sample rate (typically 44.1 kHz). Pitch is extracted via Librosa’s piptrack, which estimates fundamental frequencies and magnitudes across time frames. The dominant pitch per frame is selected based on maximum magnitude, excluding zero values. These frequencies are then converted to cents relative to 440 Hz (A4) using a logarithmic scale.
ShrutiSense processes the cents sequence along with a user-specified raga (e.g., Yaman), applying raga-specific rules to transform the pitch sequence into a Carnatic form, outputting a new cents sequence.
To convert this back to audio, each cent value is transformed into frequency, then rendered as 0.5-second stereo 16-bit PCM sine waves at 44.1 kHz. These are concatenated into a continuous stream, played in real-time via an audio mixer, and saved as a WAV file.
\section{Experimental Evaluation and Results}

To evaluate our system, we employ a comprehensive set of metrics addressing both technical accuracy and cultural authenticity. Pitch assessment includes Shruti Classification Accuracy, which measures the percentage of correctly identified Shruti states, and Average Pitch Error (APE), calculated as the mean deviation in cents between predicted and true pitches over time. Musical grammar is evaluated via Raga Grammar Compliance, which reflects the alignment of melodic transitions with raga-specific rules, and Pakad Pattern Recognition, which measures detection accuracy for characteristic melodic phrases. Computational performance is analyzed through processing time per sequence, memory usage and scalability, and real-time responsiveness.

\subsection{Correction Task Performance}

\begin{table}[h]
\centering
\caption{Pitch Correction Performance Comparison (Yaman Raga, Corruption = 0.4)}
\begin{tabular}{@{}lcccc@{}}
\toprule
\textbf{Method} & \textbf{Shruti Acc. (\%)} & \textbf{Mean Error (cents)} &  \textbf{Time (ms)} \\
\midrule
\textbf{GC-SHMM} & \textbf{84 $\pm$ 0.4} & \textbf{107.6 $\pm$ 3.6}  & \textbf{12.5 $\pm$ 0.3} \\
\textbf{Shruti FST} & \textbf{91.3 $\pm$ 0.2} & \textbf{45.6 $\pm$ 1.4} & \textbf{0.1} \\
Nearest Cent & 89.4 $\pm$ 0.3 & 51.8 $\pm$ 1.4  & 0.1 \\
Random & 12.6 $\pm$ 0.3 & 452.6 $\pm$ 2.4 & 0 \\
\bottomrule
\end{tabular}
\end{table}

Because the correction task is the most similar to a real-world example, we deemed it necessary to compare our models against some baselines and include statistical significance tests. Namely, we compare them against a basic model that corrects pitches to the nearest cent value in our 22-Shruti scale and a model that corrects pitches to a random cent value in the scale. The statistics for the correction task are displayed in Figure 1. We can clearly see that the FST beats the other models and significantly outperforms the HMM in terms of accuracy and processing speed over 900 simulations. The HMM portion of ShrutiSense achieved a mean accuracy of $84 \pm 0.4 \%$, while the FST portion achieved a mean accuracy of $91.3 \pm 0.2 \%$. Our confidence interval is $95\%$. 

The statistical analysis reveals strong and significant differences in accuracy across the models evaluated. Pairwise comparisons show that both FST and Nearest Cent outperform HMM, while all structured models far exceed the random baseline. The Cohen’s d values outputted by the evaluation code indicate large effect sizes for each contrast, especially against random, with FST showing the greatest improvement. A one-way ANOVA further confirms significant disparities among models with an exceptionally high F-statistic. Overall, the findings suggest that FST is the most accurate model, followed closely by nearest cent, while HMM lags behind but still vastly outperforms random. Thus, when users choose to use ShrutiSense to correct an audio file, the audio-file will go through the FST pipeline.

\subsection{Completion Task Performance}

\begin{table}[h]
\centering
\caption{Melodic Completion Performance by Missing Pattern}
\begin{tabular}{@{}lcccccc@{}}
\toprule
\textbf{Missing Pattern} & \multicolumn{2}{c}{\textbf{HMM}} & \multicolumn{2}{c}{\textbf{FST}} \\
& \textbf{Acc.} & \textbf{Error (cents)} & \textbf{Acc.} & \textbf{Error (cents)}  \\
\midrule
Random & 57.1 & 203.0 & 62.6 & 158.7  \\
Clustered & 40.3 & 344.9 & 26.6 & 317.0 \\
Structured & 82.9 & 48.5 & 70.5 & 228.1 \\
\bottomrule
\end{tabular}
\end{table}

As the completion task is not generally a real-world use case, we deemed it unnecessary to compare it against baseline models. The statistics for the completion task are displayed in Figure 2. Clearly, the correction task is much easier for ShrutiSense than the completion task, which is expected. Overall, the HMM performed better with a mean accuracy of $60.1 \pm 30.9 \%$ vs the FST which had a mean accuracy of $48.6 \pm 22.2 \%$. That being said, the FST actually beat the HMM for the Bhairava (0.2 corruption), the Bilawal (0.2 corruption), and the Khamaaj (0.2 corruption), showing that the FST actually does as good if not better than the HMM when corruption is low. Additionally, the completion error distribution was much less spread out for the FST than the HMM. The FST was significantly faster than the HMM, as expected because of the differences in algorithmic complexity. 

\begin{figure}[htbp]
    \centering
    \begin{minipage}[b]{0.48\linewidth}
        \centering
        \includegraphics[width=\linewidth]{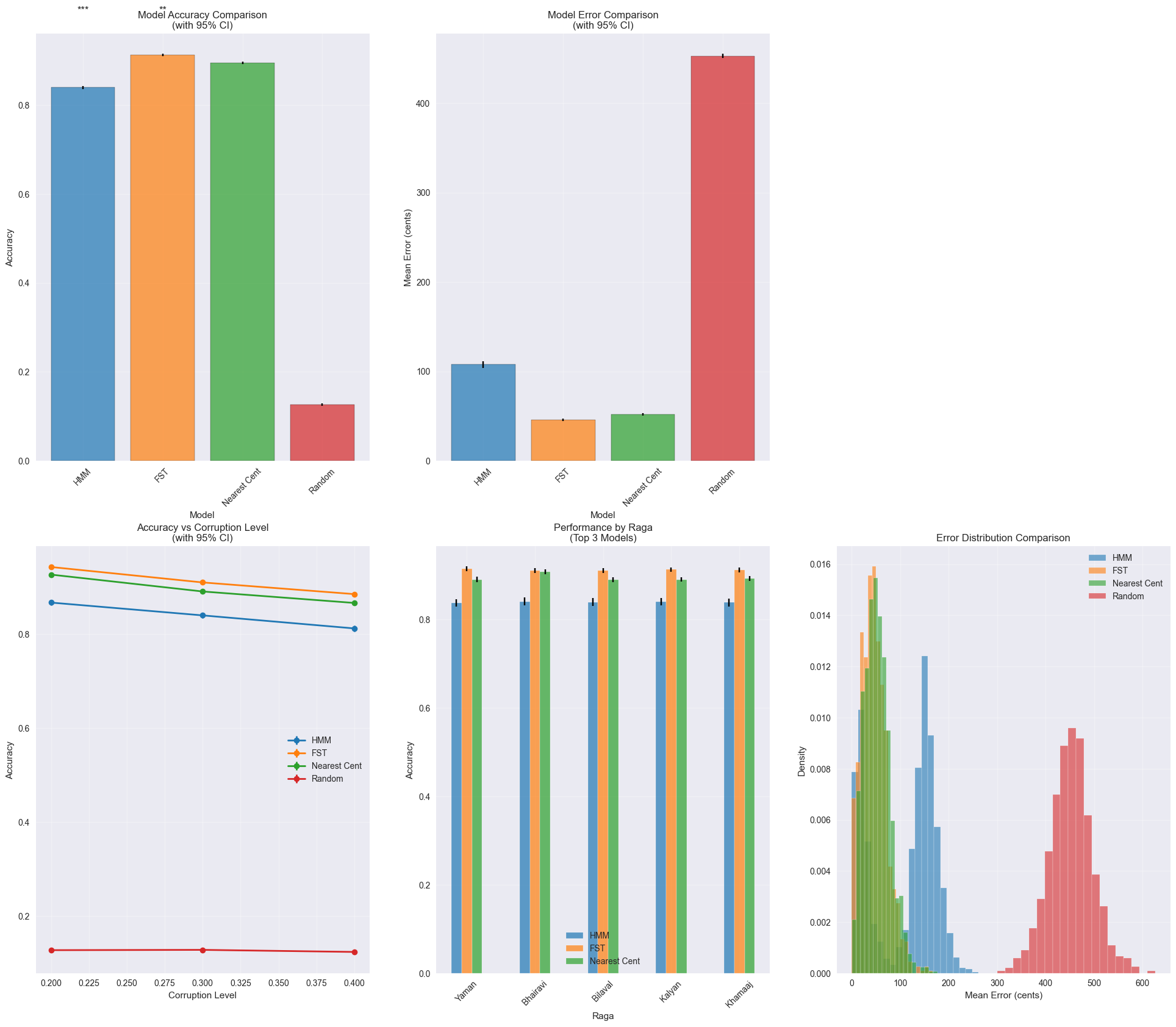}
        \caption{Correction Statistics}
        \label{fig:correction-stats}
    \end{minipage}
    \hfill
    \begin{minipage}[b]{0.48\linewidth}
        \centering
        \includegraphics[width=\linewidth]{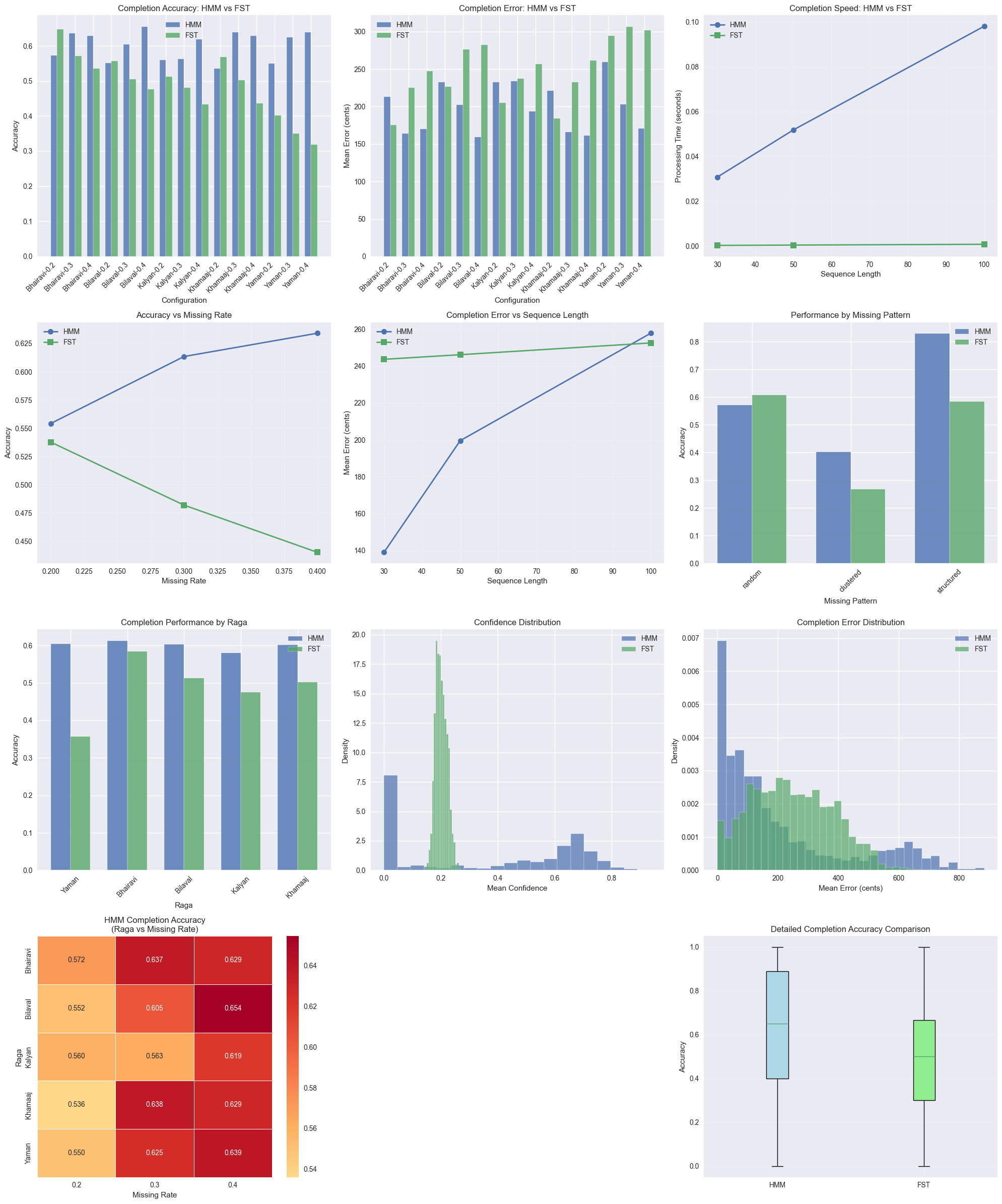}
        \caption{Completion Statistics}
        \label{fig:completion-stats}
    \end{minipage}
\end{figure}

\subsection{Robustness Analysis}

Noise resilience testing demonstrates that ShrutiSense (FST model) maintains an average accuracy of 91.3\% with input noise up to \textpm50 cents, as introduced by quantization corruption. Example sequences at corruption levels of 0.2 to 0.4 show FST accuracies ranging from 86.7\% to 90.0\%. Performance degrades gracefully at higher corruption levels, maintaining accuracy around 86.7\% at 0.4 corruption.

The system exhibits consistent performance across ragas, with minor variations: Yaman (91.1\% accuracy), Bhairavi (90.7\%), Bilaval (91.2\%), Kalyan (91.8\%), and Khamaaj (91.8\%). These results, derived from 900 simulations across sequence lengths of 30, 50, and 100, validate the generalizability of the grammar-based approach.

\section{Discussion and Future Work}

ShrutiSense highlights the theoretical advantage of encoding explicit cultural knowledge, demonstrating that grammar-constrained frameworks outperform purely statistical approaches in microtonal music processing. By incorporating domain expertise, the system maintains computational efficiency while respecting musical traditions. The complementary strengths of Hidden Markov Models (HMMs), which capture complex temporal dependencies, and Finite-State Transducers (FSTs), which enable fine-grained correction, suggest that hybrid architectures may further enhance performance. Our results have shown that for most real-life scenarios (correction task), the FST is most promising for ShrutiSense. However, in cases when users need to fill in missing notes (completion task), the HMM is most efficient. Despite promising results, current limitations include reliance on pre-defined raga grammars, assumption of monophonic input, limited modeling of ornamental nuances, and the requirement for tonic identification. Future work will focus on adaptive raga learning through unsupervised corpus analysis, end-to-end audio integration with robust pitch estimation for noisy inputs, expanded ornament modeling that treats gamakas and other microtonal inflections as first-class entities, and multi-voice extensions to handle drones and polyphonic textures common in Indian classical ensembles. We also envision cross-cultural adaptation to diverse microtonal traditions such as Carnatic music, Middle Eastern maqam systems, and contemporary non-Western composition practices.

\medskip
{
\small
\bibliographystyle{plainnat}
\nocite{*}
\bibliography{neurips_2025}


\end{document}